\documentclass[showpacs,preprintnumbers,amsmath,amssymb]{revtex4}

\usepackage{graphicx}
\usepackage{dcolumn}
\usepackage{bm}

\begin{document}

\newcommand\al{\alpha} \newcommand\be{\beta}
\newcommand\ga{\gamma} \newcommand\x{\hat x}
\newcommand\vv{\mathbf v} \newcommand\av{\mathbf c}
\newcommand\D{\mathrm D} \newcommand\G{\Gamma}
\newcommand\UU{\mathbf U} \newcommand\V{\mathbf V}
\newcommand\rr{\mathbb R}\newcommand\J{\hat J}
\newcommand\xx{\mathbf x}\newcommand\Q{\mathbf Q}
\newcommand\hh{\mathbf h}

\title{Can material time derivative be objective?}
\author{T. Matolcsi* and P. V\'an**}
\address{*E\"otv\"os Loránd University, Department of Applied Analysis \\
Budapest, Hungary\\
**(corresponding author) KFKI Research Institute for Particle and
Nuclear Physics, Department of Theoretical Physics\\
1125 Budapest 141, Pf. 49., Hungary\\
T: +36-12922222/3397 \\
F: +36-13922727}



\email{vpet@eik.bme.hu}

\date{\today}

\begin{abstract}
The concept of objectivity in classical field theories is
traditionally based on time dependent Euclidean transformations. In
this paper we treat objectivity in a four-dimensional setting,
calculate Christoffel symbols of the spacetime transformations, and
give covariant and material time derivatives. The usual objective
time derivatives are investigated.
\end{abstract}

\maketitle


\section{About objectivity}

The usual concept of objectivity in classical field theories is
based on time-dependent Euclidean transformations. The importance of
these transformations was recognized by Noll in 1958 \cite{Nol58a}
and later on became an important tool to restrict constitutive
functions through the principle of material frame-indifference (see
e.g. \cite{TruNol65b,Mul85b,Sil97b,Spe98a}).

Later on the principle of material frame-indifference was criticized
by several authors from different points of view
\cite{Mul72a,EdeMcl73a,Mur83a}. Several authors argued that some
consequences of the usual mathematical formulation of the principle
contradict the experimental observations. We emphasize that one
should make a clear distinction between the {\it principle} of
material frame-indifference (which is physically well-justified) and
its {\it mathematical formulation} (see e.g.
\cite{Rys85a,Mat86a,MatGru96a,BerSve01a,Mus98a}). There are several
different opinions in the literature concerning the mathematical
formulation  and recent research papers indicate that the discussion
does not seem to settle \cite{Mur03a,Liu03a,Mur05a,Liu05a}.

The concept of material frame-indifference is inherently related to
the notion of objectivity. In this paper, as a first step towards a
possible solution of the problems of material frame-indifference
mentioned above, we investigate the concept of objectivity.

Some well-known problems arise from the definition of objectivity
which mainly concern quantities containing derivatives. It seems to us that
the problems take their origin from the fact that objectivity is
defined for three-dimensional vectors but differentiation -- with
respect to time and space together -- results in a four-dimensional
covector. Therefore, we propose to extend the definition of objectivity to a four-dimensional setting.

We start with the usual transformation rules of Noll \cite{Nol58a}
for spacetime variables:
\begin{equation}\label{trafo}
\hat t=t, \qquad \hat{\xx}=\hh(t)+ \Q(t)\xx.
\end{equation}

Though time is not transformed, the transformation of space variables
contains time, so this is in fact a four-dimensional transformation.
Let us write in the form
\begin{equation*}
\x^0=x^0,\quad \x^\al =h^\al + Q^\al_{\ \be}x^\be;\qquad \text{in short} \qquad\x^k =\x^k(x)
\end{equation*}
where Latin indices run trough $0,1,2,3$, Greek indices run through
$1,2,3$ and the Einstein summation rule is used.

It is well-known from differential geometry (see e.g. in
\cite{ChoAta82b}) that $C$ is a four-dimensional objective vector if
it transforms according to
\begin{equation*}
\hat C^i = \J^i_jC^j
\end{equation*}
where
\begin{equation*}
\J^i_j:=\frac{\partial\x^i}{\partial x^j}
\end{equation*}
is the Jacobian matrix of the transformation.

In the present case, in a block matrix form,
\begin{equation*}
\J= \begin{pmatrix} 1 & 0 \\ \dot{\hh}
+ \dot \Q\xx  & \Q\end{pmatrix}.
\end{equation*}

Accordingly, we say that a four-vector $(C^0,\mathbf C)$ is objective if
it transforms according to the rule
\begin{equation*}
\begin{pmatrix}\hat C^0\\ \hat{\mathbf C}\end{pmatrix} =
\begin{pmatrix}
1 & 0 \\ \dot{\hh}
+ \dot \Q\xx  & \Q\end{pmatrix}
\begin{pmatrix} C^0 \\ \mathbf C\end{pmatrix},
\end{equation*}
i.e.
\begin{equation}\label{vectrafo}
\hat C^0= C^0, \qquad \hat{\mathbf C} =
(\dot{\hh} + \dot \Q\xx)C^0 + \Q\mathbf C.
\end{equation}

If the four-vector is in fact a three-vector, i.e. $C^0=0$, then
we get back the usual formula.
\begin{equation}\label{uvectrafo}
\hat{\mathbf C} = \Q\mathbf C.
\end{equation}

However, some important physical quantities are four-vectors.
Namely, consider a motion $\mathbf r$ of a mass point. Then in a
four-dimensional setting it is described by $(t,\mathbf r(t))$. Its
time derivative is $(1,\vv)$ where $\vv=\dot{\mathbf r}$. According
to \eqref{trafo} we have
 \begin{equation*}
\hat{\mathbf r}=\hh +\Q\mathbf r,
\end{equation*}
thus
\begin{equation*}
\qquad \hat{\vv} = \dot{\hh} + \dot \Q\mathbf r + \Q\dot{\mathbf r}.
\end{equation*}
As a consequence, we conclude from \eqref{vectrafo} that the four-velocity
\begin{equation}\label{absvel}
V:=(1,\vv)
\end{equation}
is an objective four-vector. The three-velocity --
i.e. the quantity $(0,\vv)$ -- is not objective.

\section{Covariant derivatives}

Since nonrelativistic spacetime is flat (has an affine
structure), it is known from differential geometry that a
distinguished covariant differentiation $\D$ is assigned to it. (In
differential geometry the covariant differentiation is usually
denoted by $\nabla$ but in spacetime this symbol usually refers to
spacelike derivatives.) This means that if $a$ is a scalar field in
spacetime, then $\D a$ is a covector field, if $C$ is a vector
field, then $\D C$ is mixed tensor field; in coordinates
\begin{equation*}
\D a \sim \D_ia = \partial_ia,
\end{equation*}
\begin{equation*}
\D C \sim \D_iC^j = \partial_iC^j +\G^i_{jk}C^k,
\end{equation*}
where  $\G^i_{jk}$ are the Christoffel symbols of the coordinatization, which
are all zero if and only if the coordinatization is linear (affine). It
is worth emphasizing: {\it the coordinates of the covariant
derivative of a vector field do not equal the partial derivatives of
the vector field if the coordinatization is not linear}. The
spacetime coordinatization is linear if and only if the underlying
observer is inertial.

If $\x$-s are inertial, linear coordinates, then the Christoffel
symbols with respect to the coordinates $x$ have the form
\begin{equation}\label{chris}
\G^i_{jk}:=\frac{\partial^2 \x^m}{\partial x^j\partial x^k}
\frac{\partial x^i}{\partial \x^m}=
-\frac{\partial^2 x^i}{\partial \x^m\partial\x^l}
\frac{\partial \x^m}{\partial x^j}
\frac{\partial \x^l}{\partial x^k}.
\end{equation}

(See the Appendix.)

A straightforward calculation yields for the coordinatization
\eqref{trafo} that
\begin{equation}\label{chris1}
\G^0_{jk}=0, \qquad \G^\al_{00}=
\bigl(\Q^{-1}(\ddot{\mathbf h} + \ddot\Q\xx)\bigr)^\al =
\bigl(\Q^{-1} \ddot{\mathbf h} + (\dot{\mathbf \Omega} +
    \mathbf{\Omega}\mathbf{\Omega})\xx)\bigr)^\al,
\end{equation}
\begin{equation}\label{chris2}
\G^\al_{0\be}=\Omega^\al_{\ \be},\qquad
\G^\al_{\ga\be}=0,
\end{equation}
where $\mathbf{\Omega}:=\Q^{-1}\dot\Q$ is the angular velocity of
the observer.

\section{Material time derivative}

Let us describe a continuum in spacetime in such a way that to each
spacetime point $x$ we assign the absolute velocity (four-velocity) $V(x)$
of the particle at $x$. Then $V$ is an objective vector field.

The flow generated by the vector field $V$ is a well-known notion in
differential geometry: $F_t(x)$ is the point at time $t$ of the
integral curve of $V$ passing through $x$ (Figure 1). Physically: an integral
curve of $V$ is the history of a particle of the continuum.
\begin{figure}[ht]
\centering
\includegraphics[height=10cm]{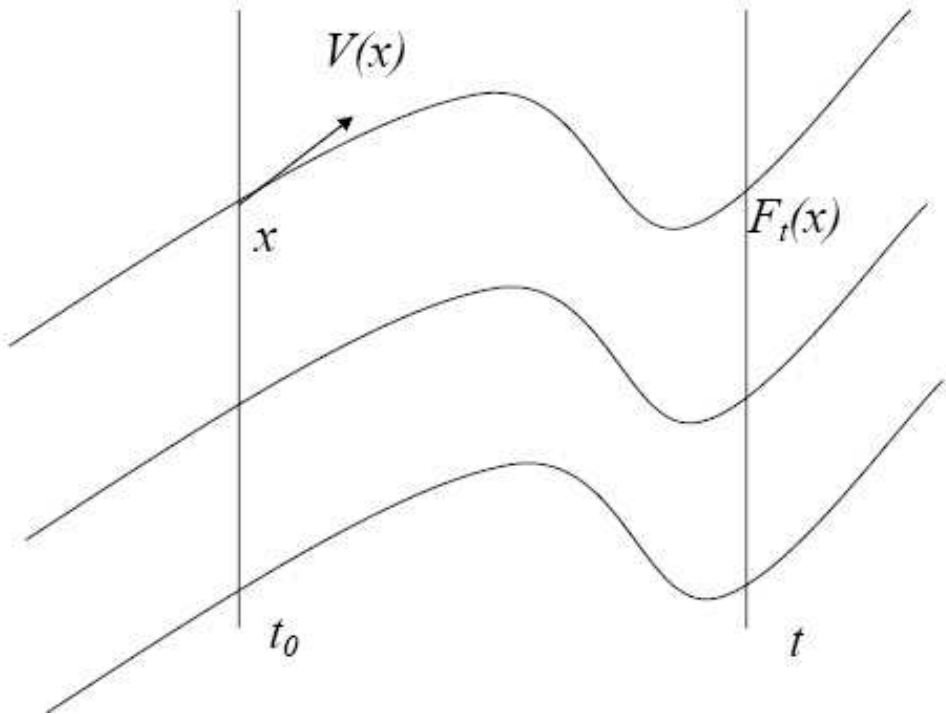}
\caption{Integral curves and flow of the vector field V.}
\label{Fig1}
\end{figure}

Let us consider a quantity $\Phi$ of any tensorial order defined in
spacetime. Then the function $t\mapsto \Phi(F_t(x))$ is the change
in time of the quantity along an integral curve i.e. at a particle
of the continuum. Then it is known from differential geometry that
\begin{equation}
\frac{d}{dt}\Phi(F_t(x)) = (\D_V\Phi )(F_t(x)),
\end{equation}
where $\D_V\Phi$ is the covariant derivative of $\Phi$ according to $V$.

We call $\D_V\Phi$ the {\it material time derivative} of $\Phi$.

Mathematically $\D_V$ is a scalar operation, i.e. the tensorial rank of
$\D_V\Phi$ equals that of $\Phi$. In coordinates:
\begin{equation}
\D_Va = V^j\D_ja = V^j\partial_ja \qquad \text{$a$ is scalar}
\end{equation}
\begin{equation}\label{ezajo}
(\D_VC)^i = V^j\D_jC^i = V^j(\partial_jC^i + \G^i_{jk}C^k)
 \qquad \text{$C$ is vector}.
\end{equation}
In view of \eqref{chris1}, \eqref{chris2} and \eqref{absvel}, we have
\begin{equation}
\D_Va = (\partial_0 + \vv\cdot\nabla)a
\end{equation}
and if $C=(0,\mathbf C)$ is a spacelike vector field, then
\begin{equation}
\D_V\mathbf C= (\partial_0 + \vv\cdot\nabla +\boldsymbol\Omega)\mathbf C.
\end{equation}
(If $C^0\neq0$, the timelike component is not zero, then a
further term containing $C^0$ enters the expression of the spacelike
component.)

The material time derivative of a vector -- even if it is spacelike
 -- is not given by $\partial_0 + \vv\cdot\nabla$.

\section{Objective time derivatives}

In usual literature $\partial_0 + \vv\cdot\nabla$ is considered to
be the material time derivation. This applied to scalars results in
scalars but applied to vectors does not result in vectors; that is
why it is always stated that this operation is not objective. Since the problem of proper objective time derivatives relates to several phenomena in physical
theories (e.g. in rheology \cite{Old49a,Ver97b}),  `objective time derivatives' are looked for in such a way that the above operation is supplemented by some terms which formally do not refer to the observer \cite{BamMor80a}.
This means in our formalism that one looks for an objective expression, containing $\partial_0 + \vv\cdot\nabla$, in which the Christoffel symbols do not appear. For instance, let us take the objective quantity

\begin{equation}\label{kieg}
C^j\D_j V^i= C^j\partial_j V^i + C^j\G^i_{jk}V^k
\end{equation}
Evidently, the difference of the true material time derivative
\eqref{ezajo} and the above expression is objective as well (being the difference of two objective quantities). Since
$\G$ is symmetric in its lower indices, we obtain that
\begin{equation*}
V^j\D_j C^i - C^j\D_j V^i= V^j\partial_j C^i - C^j\partial_j V^i.
\end{equation*}
The right-hand side does not contain the Christoffel symbols, it is
given by partial derivatives only, nevertheless it is objective. For
a spacelike vector $(0,\mathbf C)$ the right-hand side can be
written in the form
\begin{equation}
(\partial_0 +\vv\cdot\nabla)\mathbf C - \mathbf L\cdot\mathbf C,
\label{upder}\end{equation}
where $\mathbf L$ is the velocity gradient. This expression is just
the `upper convected derivative' of $\mathbf C$. In a similar way we
get the lower convected derivative and the Jaumann derivative as well.

\section{Conclusions}

\begin{enumerate}

\item We propose that objectivity be extended to a
four-dimensional setting.

\item The four-dimensional covariant differentiation is a fundamental
fact of nonrelativistic spacetime. {\it The coordinates of the
covariant derivative of a vector field do not equal the partial
derivatives of the vector field if the coordinatization is not
linear; the Christoffel symbols enter}. The essential part of the
Christoffel symbols is the angular velocity of the observer.

\item Usual treatments leave the covariant differentiation out of
consideration; they involve only partial derivatives which, of
course, are not objective. A number of problems arise from this
fact.

\item Material time derivative of a quantity, in a physically proper
sense, ought to be defined by the time derivative of the quantity along the
particles of a continuum.

\item The mathematical expression of material time differentiation
concerns the covariant differentiation, so its form relative to an
observer is not given by partial derivatives only, the Christoffel
symbols (the angular velocity of the observer) enter; for spacelike
vectors the material time differentiation has the form
\begin{equation*}
\partial_0 + \vv\cdot\nabla + \boldsymbol\Omega .
\end{equation*}

\item In the literature Jaumann derivative, upper and lower convected derivatives are introduced because instead of the
above true material time derivative authors look for objective
expressions containing $\partial_0 + \vv\cdot\nabla$ and partial
derivatives only.
\end{enumerate}

\section{Acknowledgements}

This research was supported by OTKA T048489.

\section{Appendix}

Let $\x^j$ denote inertial (i.e. affine) coordinates of spacetime
and let $x^i$ be arbitrary coordinates. Then
\begin{equation}\label{jacobi}
J(x)^i_j:=\frac{\partial x^i}{\partial\x^j},\qquad
\J(\x)^j_i:=\frac{\partial\x^j}{\partial x^i}
\end{equation}
and we know that $J(\x(x))^i_l \J(x))^l_k=\delta^i_k$ from which it
follows that
\begin{equation}\label{jjder}
0 =
\frac{\partial^2 x^i}{\partial \x^m\partial \x^l}
\frac{\partial \x^m}{\partial x^j}
\frac{\partial \x^l}{\partial x^k} +
\frac{\partial x^i}{\partial \x^l}
\frac{\partial^2 \x^l}{\partial x^j\partial x^k}
\end{equation}

For the coordinates of a vector field $C$ we have
\begin{equation}\label{vekkor}
C^i(x) = J^i_l(\x(x))\hat C^l(\x(x)), \qquad
\hat C^l(\x)=\J^l_k(x(\x)) C^k(x(\x)).
\end{equation}

The covariant derivative
$\D C$ of a vector field is a tensor field whose coordinates in affine
coordinatization are just the partial derivatives of the vector field:
$(\D \hat C)^l_m=\D_m\hat C^l=\frac{\partial \hat C^l}{\partial \x^m}$.
According to the transformation of tensors, the arbitrary coordinates
of this tensor field are
\begin{equation}\label{kovdr}
(\D_j C^i)(x) = J^i_l(\x(x))(\D \hat C)^l_m(\x(x)))\J^m_j(x)
\end{equation}
For the partial derivatives of $C^i$, from \eqref{vekkor} we infer
(with a loose notation, omitting the variables)
\begin{equation*}
\frac{\partial C^i}{\partial x^j}
 =\frac{\partial J^i_l}{\partial \x^m}
\frac{\partial \x^m}{\partial x^j}\hat C^l
+ J^i_l\frac{\partial \hat C^l}{\partial \x^m}
\frac{\partial \x^m}{\partial x^j}.
\end{equation*}

Then with the aid of \eqref{jacobi}, \eqref{kovdr} and \eqref{vekkor} we obtain
\begin{equation*}
\D_j C^i = \frac{\partial C^i}{\partial x^j} +
\G^i_{jk}C^k.
\end{equation*}
Here we have got the first form of the Christoffel symbol given in
\eqref{chris}; the second form comes from \eqref{jjder}.

2. The inverse of the transformation \eqref{trafo} is
$$ t=\hat t, \qquad \xx=\Q(\hat t)^{-1}(\hat{\xx} - \hh(\hat t)).$$

Then
\begin{equation}\label{jjacob}
J= \begin{pmatrix} 1 & 0 \\ -\Q^{-1}\dot \Q\Q^{-1}
(\hat{\xx} - \hh) - \Q^{-1}\dot{\hh}  & \Q^{-1}\end{pmatrix}.
\end{equation}

\end{document}